\documentclass[11pt,a4paper]{article}
\usepackage{jheppub}
\usepackage{stmaryrd}

\title{The secret symmetries of the AdS/CFT reflection matrices}

\author[a,b]{Vidas Regelskis}

\affiliation[a]{Department of Mathematics, University of York,\\Heslington, York YO10 5DD, UK}
\affiliation[b]{Institute of Theoretical Physics and Astronomy of Vilnius University,\\Go\v{s}tauto 12, Vilnius 01108, Lithuania}

\emailAdd{vr509@york.ac.uk}

\abstract{We find new twisted Yangian symmetries of the AdS/CFT 
reflection matrices for the Y=0 maximal giant graviton and D5-brane. 
These new symmetries originate from the known secret symmetries of 
the Yangian symmetry of the AdS/CFT S-matrix.}

\begin{document}

\maketitle


\section{Introduction}

The recent progress in solving the planar limit of AdS/CFT was accelerated 
a lot by underlying integrable structures (see review \cite{review} and references therein).
Some problems that look highly challenging in the perturbative regime,
surrender uncompromisingly to the power of underlying symmetries in
the planar limit. The striking success in finding the light-cone worldsheet 
scattering $S$-matrices and reflection $K$-matrices from various boundaries was achieved due
to underlying $\mathfrak{psu}(2,2|4)$ symmetry which in the scattering theory may be 
factorized into two copies, left and right, of the centrally-extended $\mathfrak{su}(2|2)$
algebra and its Yangian extension \cite{Beisert2,Beisert4,AT1,AT2}. 

The Yangian of the centrally-extended $\mathfrak{su}(2|2)$ is an
infinite-dimensional algebra generated by the charges $\mathbb{J}^A$, $\widehat{\mathbb{J}}^A$,
whose co-products are given by expressions
\begin{align}
\Delta\mathbb{J}^{A} & =\mathbb{J}^{A}\otimes1+1\otimes\mathbb{J}^{A},
 \qquad\Delta\widehat{\mathbb{J}}^{A} = \widehat{\mathbb{J}}^{A}\otimes1+1\otimes\widehat{\mathbb{J}}^{A}+\frac{1}{2}f_{\; BC}^{A}\,\mathbb{J}^{B}\otimes\mathbb{J}^{C},
\end{align}
of Drinfeld's first realization \cite{Drinfeld}. Invariance
under the Yangian charges fixes the $S$-matrix up to an overall
phase which can further be constrained by the crossing symmetry \cite{Janik}.

The $\mathfrak{su}(2|2)$ algebra has a $\mathfrak{u}(1)$ outer automorphism
extending the algebra to $\mathfrak{u}(2|2)$. However, the
additional $\mathfrak{u}(1)$ charge, which acts as a helicity operator,
\begin{equation}
\mathbb{B}\left|\phi_{a}\right\rangle =+I\left|\phi_{a}\right\rangle ,\qquad\mathbb{B}\left|\psi_{\alpha}\right\rangle 
  =-I\left|\psi_{\alpha}\right\rangle ,\label{B}
\end{equation}
is not a symmetry of the $S$-matrix and the eigenvalue $I$ 
is not constrained to any particular value\footnote{The charge $\mathbb B$ 
can not be added to the centrally-extended $\mathfrak{su}(2|2)$ algebra as 
it would generate infinite series of new charges that do not close under the 
Lie brackets thus do not constitute a Lie algebra. However it can be related 
to the algebra by some non-linear commutation relations \cite{dLthesis}.}. 
Strikingly, this charge has a Yangian partner $\widehat{\mathbb{B}}$ which is 
known as a `secret symmetry' of the $S$-matrix and is the same for left and 
right factors \cite{Torrielli4}. Interestingly, the non-trivial part of 
the co-product of $\widehat{\mathbb{B}}$ appears to be the same
as the $\varepsilon$-correction of $\Delta\widehat{\mathbb{H}}$ in the
limit $\varepsilon\rightarrow0$ of the exceptional superalgebra $\mathfrak{d}(2,1;\varepsilon)$
\cite{Matsumoto1}. Furthermore, it generates new secret symmetries
of the $S$-matrix, which have no Lie algebra analog. These hints
suggest that the secret symmetries may be regarded as the remnants
of some higher, yet-to-be-discovered, symmetry.

Strong support for the last statement also comes from the analysis
of the scattering amplitudes of $\mathcal{N}=4$ SYM theory in the
planar limit. It is known that the planar tree-level amplitudes not only
enjoy the Yangian symmetry of the superconformal algebra $\mathfrak{psu}(2,2|4)$,
but also respect an additional symmetry, which is referred as a `bonus
symmetry' \cite{Bonus}. This additional symmetry, $\widehat{\mathfrak{B}}$,
is the Yangian partner of the $\mathfrak{u}(1)$ outer automorphism
$\mathfrak{B}$ of $\mathfrak{psu}(2,2|4)$ first observed in \cite{Intriligator}. 
Similarly as for the AdS/CFT worldsheet $S$-matrix, the helicity operator $\mathfrak{B}$
is not a symmetry of the planar $\mathcal{N}=4$ SYM $S$-matrix,
while $\widehat{\mathfrak{B}}$ is, - at least for the tree-level MHV
(\textit{Maximum Helicity Violating}) scattering amplitudes.

Secret symmetry also appears to play a role in the quantum affine algebra 
$\widehat{\mathcal{Q}}$ based on the centrally-extended $\mathfrak{sl}(2|2)$, 
where it manifests itself explicitly when taking the Yangian limit \cite{NGM}.

In this letter we want to explore the secret symmetries of the twisted
boundary Yangians describing the reflection from the $Y=0$ maximal 
giant graviton \cite{AN,MR2} and $D5$-brane \cite{MR3}. In section
2 we briefly review the secret symmetries of the $S$-matrix.
In section 3, based on the considerations on the $S$-matrix in the section
2, we build the secret symmetries hiding inside the twisted boundary Yangians. 


\section{Secret symmetries of the S-matrix}

It was shown in \cite{Torrielli4} and confirmed in \cite{BS} that the additional charge
\begin{equation}
\Delta\widehat{\mathbb{B}} = \widehat{\mathbb{B}}\otimes1+1\otimes\widehat{\mathbb{B}}
  - \frac{1}{2}(\mathbb{Q}_{\alpha}^{\enskip a}\otimes\mathbb{G}_{a}^{\enskip\alpha}
  + \mathbb{G}_{a}^{\enskip\alpha}\otimes\mathbb{Q}_{\alpha}^{\enskip a}),\label{secretB}
\end{equation}
where $\widehat{\mathbb{B}}$ is the level-1 partner of (\ref{B}) with
the eigenvalue \begin{equation}
\widehat{I}=\frac{ig}{8}\Bigl(x^{+}-\frac{1}{x^{+}}+x^{-}-\frac{1}{x^{-}}\Bigr),\end{equation}
is a symmetry of the $S$-matrix.
Furthermore, this novel symmetry generates several 
new secret symmetries of the $S$-matrix that do not have a Lie algebra
analog. They were originally found by computing the commutators 
$[\Delta\widehat{\mathbb{B}},\Delta\mathbb{Q}_{a}^{\enskip\alpha}]$
and $[\Delta\widehat{\mathbb{B}},\Delta\mathbb{G}_{\alpha}^{\enskip a}]$
and taking linear combinations with the Yangian charges $\Delta\widehat{\mathbb{Q}}_{a}^{\enskip\alpha}$
and $\Delta\widehat{\mathbb{G}}_{\alpha}^{\enskip a}$ (see \cite{Torrielli4} for the details).

These new secret symmetries generated by (\ref{secretB}) are
\begin{align}
\Delta\mathbb{Q}_{\alpha,+1}^{\enskip a} & =\mathbb{Q}_{\alpha,+1}^{\enskip a}\otimes1+1\otimes\mathbb{Q}_{a,+1}^{\enskip\alpha}-\frac{1}{2}\mathbb{L}_{\alpha}^{\enskip\gamma}\otimes\mathbb{Q}_{\gamma}^{\enskip a}+\frac{1}{2}\mathbb{Q}_{\gamma}^{\enskip a}\otimes\mathbb{L}_{\alpha}^{\enskip\gamma}\nonumber \\
 & \qquad-\frac{1}{2}\mathbb{R}_{c}^{\enskip a}\otimes\mathbb{Q}_{\alpha}^{\enskip c}+\frac{1}{2}\mathbb{Q}_{\alpha}^{\enskip c}\otimes\mathbb{R}_{c}^{\enskip a}-\frac{1}{4}\mathbb{H}\otimes\mathbb{Q}_{\alpha}^{\enskip a}+\frac{1}{4}\mathbb{Q}_{\alpha}^{\enskip a}\otimes\mathbb{H},\nonumber \\
\Delta\mathbb{Q}_{\alpha,-1}^{\enskip a} & =\mathbb{Q}_{\alpha,-1}^{\enskip a}\otimes1+1\otimes\mathbb{Q}_{a,-1}^{\enskip\alpha}-\frac{1}{2}\varepsilon_{\alpha\gamma}\,\varepsilon^{ac}\,\mathbb{G}_{c}^{\enskip\gamma}\otimes\mathbb{C}+\frac{1}{2}\varepsilon_{\alpha\gamma}\,\varepsilon^{ac}\,\mathbb{C}\otimes\mathbb{G}_{c}^{\enskip\gamma},\nonumber \\
\Delta\mathbb{G}_{a,+1}^{\enskip\alpha} & =\mathbb{G}_{a,+1}^{\enskip\alpha}\otimes1+1\otimes\mathbb{G}_{a,+1}^{\enskip\alpha}+\frac{1}{2}\mathbb{L}_{\gamma}^{\enskip\alpha}\otimes\mathbb{G}_{a}^{\enskip\gamma}-\frac{1}{2}\mathbb{G}_{a}^{\enskip\gamma}\otimes\mathbb{L}_{\gamma}^{\enskip\alpha}\nonumber \\
 & \qquad+\frac{1}{2}\mathbb{R}_{a}^{\enskip c}\otimes\mathbb{G}_{c}^{\enskip\alpha}-\frac{1}{2}\mathbb{G}_{c}^{\enskip\alpha}\otimes\mathbb{R}_{a}^{\enskip c}+\frac{1}{4}\mathbb{H}\otimes\mathbb{G}_{a}^{\enskip\alpha}-\frac{1}{4}\mathbb{G}_{a}^{\enskip\alpha}\otimes\mathbb{H},\nonumber \\
\Delta\mathbb{G}_{a,-1}^{\enskip\alpha} & =\mathbb{G}_{a,-1}^{\enskip\alpha}\otimes1+1\otimes\mathbb{G}_{a,-1}^{\enskip\alpha}+\frac{1}{2}\varepsilon_{ac}\,\varepsilon^{\alpha\gamma}\,\mathbb{Q}_{\gamma}^{\enskip c}\otimes\mathbb{C}^{\dagger}-\frac{1}{2}\varepsilon_{ac}\,\varepsilon^{\alpha\gamma}\,\mathbb{C}^{\dagger}\otimes\mathbb{Q}_{\gamma}^{\enskip c},\label{secretQG}
\end{align}
where the new secret charges $\mathbb{Q}_{\alpha,\pm1}^{\enskip a}$
and $\mathbb{G}_{a,\pm1}^{\enskip\alpha}$ are defined as 
\begin{align}
\mathbb{Q}_{\alpha,+1}^{\enskip a} & =i\frac{g}{2}\,\mathbb{Q}_{\alpha}^{\enskip a}\,(u\Pi_{b}+v\Pi_{f}), &  & \mathbb{Q}_{\alpha,-1}^{\enskip a}=i\frac{g}{2}\,\mathbb{Q}_{\alpha}^{\enskip a}\,(v\Pi_{b}+u\Pi_{f}),\nonumber \\
\mathbb{G}_{a,+1}^{\enskip\alpha} & =i\frac{g}{2}\,\mathbb{G}_{a}^{\enskip\alpha}\,(v\Pi_{b}+u\Pi_{f}), &  & \mathbb{G}_{a,-1}^{\enskip\alpha}=i\frac{g}{2}\,\mathbb{G}_{a}^{\enskip\alpha}\,(u\Pi_{b}+v\Pi_{f}),
\end{align}
with $\Pi_{b}$ and $\Pi_{f}$ being the projectors onto bosons and
fermions respectively and 
\begin{equation}
u=\frac{1}{2}(x^{+}+x^{-}),\qquad v=\frac{1}{2}\left(\frac{1}{x^{+}}+\frac{1}{x^{-}}\right).
\end{equation}
The charges of such form were first considered in constructing the classical 
r-matrix of AdS/CFT \cite{MT}.
It is easy to convince ourselves that these charges do not have Lie algebra analog, 
because the naive Lie algebra limit ({i.e.\ $u\rightarrow1$, $v\rightarrow1$) 
leads to the usual Lie algebra supercharges 
$\mathbb{Q}_{\alpha}^{\enskip a}$ and $\mathbb{G}_{a}^{\enskip\alpha}$.

We have checked that all these symmetries are respected by higher order
(two-magnon bound-state) $S$-matrices as well.


\section{Secret symmetries of the K-matrices}

In a recent series of works \cite{MR3,Maldacena2,CY,MR1,CRY} reflection
$K$-matrices were found for open strings ending on $D3$, $D5$ and $D7$
branes. It was also shown, that some of them respect not only Lie,
but also a wider, twisted boundary Yangian symmetry \cite{AN,MR2,MR3,Palla1} 
(see also \cite{MacKay1,MacKay2})
originating from the Yangian of the $S$-matrix \cite{Beisert4}. An
interesting question is if the secret symmetries (\ref{secretB})
and (\ref{secretQG}) manifest themselves in the twisted Yangians.
And indeed we find them to be present.

We will not present the explicit calculations of the invariance conditions
for the secret symmetries we will construct as they are quite straightforward 
and not very illuminating, but at the same time involve very large computer 
algebra calculations that we have performed with 
{\tt Mathematica}.\footnote{For the explicit calculations we are using the superspace formalism introduced in \cite{AF}, in which
the secret charge \eqref{secretB} is defined as
$\widehat{\mathbb{B}} = \widehat{I}\, \left( w_a \frac{\partial}{\partial w_a} - \theta_\alpha \frac{\partial}{\partial \theta_\alpha}\right)$ 
and is equivalent (up to a prefactor) to the charge $\Sigma$ introduced in \cite{MdL}.}

\paragraph{Maximal giant graviton.} The $Y=0$ maximal giant graviton
preserves a subalgebra $\mathfrak{h}=\mathfrak{su}(2|1)_{L} = \{\mathbb{L}_{\alpha}^{\enskip\beta},\;\mathbb{R}_{1}^{\enskip1},\;\mathbb{R}_{2}^{\enskip2},\;\mathbb{Q}_{\alpha}^{\enskip1},\;\mathbb{G}_{1}^{\enskip\alpha},\;\mathbb{H}\}$
of the bulk algebra $\mathfrak{g}=\mathfrak{psu}(2|2)\ltimes\mathbb{R}^{3}$.
The boundary Yangian symmetry is generated by the twisted charges
\begin{equation}
\widetilde{\mathbb{J}}^{p}:=\widehat{\mathbb{J}}^{p} + 
 \frac{1}{4}f_{\; qi}^{p}\,(\mathbb{J}^{q}\,\mathbb{J}^{i}+\mathbb{J}^{i}\,\mathbb{J}^{q}),\label{twist}
\end{equation}
the co-products of which are of the form
\begin{eqnarray}
\Delta\widetilde{\mathbb{J}}^{p} & = & \widetilde{\mathbb{J}}^{p}\otimes1+1\otimes\widetilde{\mathbb{J}}^{p}+f_{\; qi}^{p}\,\mathbb{J}^{q}\otimes\mathbb{J}^{i},\label{Y_gm}
\end{eqnarray}
where $\mathbb{J}^{i}\in\mathfrak{h}$ and $\mathbb{J}^{p\,(q)}\in\mathfrak{m}
 =\{\mathbb{R}_{1}^{\enskip2},\;\mathbb{R}_{2}^{\enskip1},\;\mathbb{Q}_{\gamma}^{\enskip2},\;\mathbb{G}_{2}^{\enskip\gamma},\;\mathbb{C},\;\mathbb{C}^{\dagger}\}$
are the generators of the subset $\mathfrak{m}=\mathfrak{g}\backslash \mathfrak{h}$. 
The boundary is a singlet in the scattering theory, thus only terms
of the form $\widetilde{\mathbb{J}}^{p}\otimes1$ in (\ref{Y_gm}) need
to be considered (see \cite{MR2} for details).

The fundamental reflection matrix, describing the scattering of fundamental
magnons from the boundary, is diagonal and the helicity generator
$\mathbb{B}$ (\ref{B}) is a symmetry of it, but the twisted (\ref{twist})
partner of the secret Yangian charge (\ref{secretB}) is not. Higher
order reflection matrices are of non-diagonal form and do not respect
either additional symmetry $\mathbb{B}$, or the twisted partner
of $\widehat{\mathbb{B}}$. 

The next step is to check if the twisted (\ref{twist}) partners of 
the additional secret charges (\ref{secretQG}) are symmetries of 
the reflection matrix. By performing the twist (\ref{twist}) we found the 
new additional twisted secret charges to be 
\begin{align}
\Delta\widetilde{\mathbb{Q}}_{\alpha,+1}^{\enskip2} & =\left(\mathbb{Q}_{\alpha,+1}^{\enskip2}+\frac{1}{2}\mathbb{Q}_{\alpha}^{\enskip2}\,\mathbb{R}_{2}^{\enskip2}-\frac{1}{2}\mathbb{R}_{1}^{\enskip2}\,\mathbb{Q}_{\alpha}^{\enskip1}+\frac{1}{2}\mathbb{Q}_{\gamma}^{\enskip2}\,\mathbb{L}_{\alpha}^{\enskip\gamma}+\frac{1}{4}\mathbb{Q}_{\alpha}^{\enskip2}\,\mathbb{H}\right)\otimes1,\nonumber \\
\Delta\widetilde{\mathbb{Q}}_{\alpha,-1}^{\enskip2} & =\left(\mathbb{Q}_{\alpha,-1}^{\enskip2}-\frac{1}{2}\varepsilon_{\alpha\gamma}\,\mathbb{C}\,\mathbb{G}_{1}^{\enskip\gamma}\right)\otimes1,\nonumber \\
\Delta\widetilde{\mathbb{G}}_{2,+1}^{\enskip\alpha} & =\left(\mathbb{G}_{2,+1}^{\enskip\alpha}-\frac{1}{2}\mathbb{G}_{2}^{\enskip\alpha}\,\mathbb{R}_{2}^{\enskip2}+\frac{1}{2}\mathbb{R}_{2}^{\enskip1}\,\mathbb{G}_{1}^{\enskip\alpha}-\frac{1}{2}\mathbb{G}_{2}^{\enskip\gamma}\,\mathbb{L}_{\gamma}^{\enskip\alpha}-\frac{1}{4}\mathbb{G}_{2}^{\enskip\alpha}\,\mathbb{H}\right)\otimes1,\nonumber \\
\Delta\widetilde{\mathbb{G}}_{2,-1}^{\enskip\alpha} & =\left(\mathbb{G}_{2,-1}^{\enskip\alpha}+\frac{1}{2}\varepsilon^{\alpha\gamma}\,\mathbb{C}^{\dagger}\,\mathbb{Q}_{\gamma}^{\enskip1}\right)\otimes1,
\end{align}
and checked that they commute with the fundamental and two-magnon bound-state 
reflection matrices. Hence they are the symmetries of the reflection matrix. 
(Reflection matrices for general Q-magnon bound-states were constructed in 
\cite{Palla1} thus the invariance condition could be easily checked.)\\

The mirror model of the $Y=0$ maximal giant graviton \cite{MR2,Palla1} preserves the
subalgebra $\mathfrak{h}=\mathfrak{su}(2|1)_{R} = \{\mathbb{R}_{a}^{\enskip b},\;
\mathbb{L}_{3}^{\enskip3},\;\mathbb{L}_{4}^{\enskip4},\;\mathbb{Q}_{3}^{\enskip a},\;\mathbb{G}_{a}^{\enskip3}\;\mathbb{H}\}$
and the complementary subset is 
$\mathfrak{m} = \{\mathbb{L}_{3}^{\enskip4},\;\mathbb{L}_{4}^{\enskip3},\;
  \mathbb{Q}_{4}^{\enskip a},\;\mathbb{G}_{a}^{\enskip4},\;\mathbb{C},\;\mathbb{C}^{\dagger}\}$.
The boundary is a singlet and the reflection matrices are diagonal
at all orders of the bound-state number; thus $\mathbb{B}$ (\ref{B}) is
a symmetry at all orders, but the twisted partner of the secret Yangian
charge (\ref{secretB}) is not (at any order). Similarly to the
previous case we have checked that the twisted partners 
\begin{align}
\Delta\widetilde{\mathbb{Q}}_{4,+1}^{\enskip a} & =\left(\mathbb{Q}_{4,+1}^{\enskip a}+\frac{1}{2}\mathbb{Q}_{4}^{\enskip c}\,\mathbb{R}_{c}^{\enskip a}+\frac{1}{2}\mathbb{Q}_{4}^{\enskip a}\,\mathbb{L}_{4}^{\enskip4}-\frac{1}{2}\mathbb{L}_{4}^{\enskip3}\,\mathbb{Q}_{3}^{\enskip a}+\frac{1}{4}\mathbb{Q}_{4}^{\enskip a}\,\mathbb{H}\right)\otimes1,\nonumber \\
\Delta\widetilde{\mathbb{Q}}_{4,-1}^{\enskip a} & =\left(\mathbb{Q}_{4,-1}^{\enskip a}-\frac{1}{2}\varepsilon^{ad}\,\mathbb{C}\,\mathbb{G}_{d}^{\enskip3}\right)\otimes1,\nonumber \\
\Delta\widetilde{\mathbb{G}}_{a,+1}^{\enskip4} & =\left(\mathbb{G}_{a,+1}^{\enskip4}-\frac{1}{2}\mathbb{G}_{c}^{\enskip4}\,\mathbb{R}_{a}^{\enskip c}-\frac{1}{2}\mathbb{G}_{a}^{\enskip4}\,\mathbb{L}_{4}^{\enskip4}+\frac{1}{2}\mathbb{L}_{3}^{\enskip4}\,\mathbb{G}_{a}^{\enskip3}-\frac{1}{4}\mathbb{G}_{a}^{\enskip4},\mathbb{H}\right)\otimes1,\nonumber \\
\Delta\widetilde{\mathbb{G}}_{a,-1}^{\enskip4} & =\left(\mathbb{G}_{a,-1}^{\enskip4}+\frac{1}{2}\varepsilon_{ac}\,\mathbb{C}^{\dagger}\,\mathbb{Q}_{3}^{\enskip c}\right)\otimes1,
\end{align}
of the secret charges (\ref{secretQG}) are symmetries of the
reflection matrix.

\paragraph{D5-brane.} The $D5$-brane preserves a diagonal subalgebra
$\mathfrak{psu}(2|2)_{+}\ltimes\mathbb{R}^{3}$ of the bulk algebra
$\mathfrak{psu}(2|2)\times\widetilde{\mathfrak{psu}}(2|2)\ltimes\mathbb{R}^{3}$
generated by \cite{CY,CRY} 
\begin{align}
\mathbb{L}_{\check{\alpha}}^{\enskip\check{\beta}} & =\mathbb{L}_{\alpha}^{\enskip\beta}+\mathbb{L}_{\bar{\dot{\alpha}}}^{\enskip\bar{\dot{\beta}}}, &  & \mathbb{Q}_{\check{\alpha}}^{\enskip\check{a}}=\mathbb{Q}_{\alpha}^{\enskip a}+\kappa\,\mathbb{Q}_{\bar{\dot{\alpha}}}^{\enskip\dot{a}},\nonumber \\
\mathbb{R}_{\check{a}}^{\enskip\check{b}} & =\mathbb{R}_{a}^{\enskip b}+\mathbb{R}_{\dot{a}}^{\enskip\dot{b}}, &  & \mathbb{G}_{\check{a}}^{\enskip\check{\alpha}}=\mathbb{G}_{a}^{\enskip\alpha}+\kappa^{-1}\mathbb{G}_{\dot{a}}^{\enskip\bar{\dot{\alpha}}}.\label{diag_symm}
\end{align}
where $\kappa^2=\pm1$; the notation for the dotted and checked indices is the same
as for undotted ones, $\dot{a},\,\check{a},\,\dot{b},\,\check{b}=\dot{1},\,\dot{2}$
and $\dot{\alpha},\,\check{\alpha},\,\dot{\beta},\,\check{\beta}=\dot{3},\,\dot{4}$;
the bar above the dotted indices acts as $\bar{\dot{3}}=\dot{4}$
and $\bar{\dot{4}}=\dot{3}$. The generators with the undotted indices
generate `left' $\mathfrak{psu}(2|2)$ and the generators with the dotted
indices generate `right' $\widetilde{\mathfrak{psu}}(2|2)$. The complementary
charges are defined as 
\begin{align}
\overline{\mathbb{L}}_{\check{\alpha}}^{\enskip\check{\beta}} & = \mathbb{L}_{\alpha}^{\enskip\beta}-\mathbb{L}_{\bar{\dot{\alpha}}}^{\enskip\bar{\dot{\beta}}}, &  & \overline{\mathbb{Q}}_{\check{\alpha}}^{\enskip\check{a}}=\mathbb{Q}_{\alpha}^{\enskip a}-\kappa\,\mathbb{Q}_{\bar{\dot{\alpha}}}^{\enskip\dot{a}},\nonumber \\
\overline{\mathbb{R}}_{\check{a}}^{\enskip\check{b}} & = \mathbb{R}_{a}^{\enskip b}-\mathbb{R}_{\dot{a}}^{\enskip\dot{b}}, &  & \overline{\mathbb{G}}_{\check{a}}^{\enskip\check{\alpha}}=\mathbb{G}_{a}^{\enskip\alpha}-\kappa^{-1}\mathbb{G}_{\dot{a}}^{\enskip\bar{\dot{\alpha}}},\label{comp_diag_symm}
\end{align}
and in the contrast to \eqref{diag_symm} annihilate the boundary by definition (see \cite{MR3} for details).

The Yangian symmetry of the $D5$-brane is generated by the twisted charges
\begin{equation}
\widetilde{\mathbb{J}}^{\check{A}} := \widehat{\overline{\mathbb{J}}}{}^{\check{A}}+\frac{1}{8}f_{\;\check{B}\check{C}}^{\check{A}}
  \Bigl(\overline{\mathbb{J}}{}^{\check{B}}\,\mathbb{J}^{\check{C}} 
   + \mathbb{J}^{\check{C}}\,\overline{\mathbb{J}}{}^{\check{B}}\Bigr),\label{twist_d}
\end{equation}
where the indices $\check{A},\;\check{B},\;\check{C},$ run through all possible charges.
The co-products of (\ref{twist_d}) acquire the form
\begin{eqnarray}
\Delta\widetilde{\mathbb{J}}^{A} & = & {\widetilde{\mathbb{J}}}^{\check{A}}\otimes1+1\otimes{\widetilde{\mathbb{J}}}^{\check{A}}+\frac{1}{2}f_{\; BC}^{A}\,\overline{\mathbb{J}}{}^{\check{B}}\otimes\mathbb{J}^{\check{C}}.
\end{eqnarray}
Based on this construction it is easy to see that the twisted (\ref{twist_d}) partner for the $D5$-brane of the secret charge (\ref{secretB}) is
\begin{equation}
\Delta\widetilde{\mathbb{B}}=\widetilde{\mathbb{B}}\otimes1-1\otimes\widetilde{\mathbb{B}}-\frac{1}{2}\bigl(\overline{\mathbb{Q}}_{\check{\alpha}}^{\enskip\check{a}}\otimes\mathbb{G}_{\check{a}}^{\enskip\check{\alpha}}+\overline{\mathbb{G}}_{\check{a}}^{\enskip\check{\alpha}}\otimes\mathbb{Q}_{\check{\alpha}}^{\enskip\check{a}}\bigr),\label{B_D5}\end{equation}
while the twisted partners of (\ref{secretQG}) are
\begin{align}
\Delta\widetilde{\mathbb{Q}}_{\check{\alpha},+1}^{\enskip\check{a}} & =\widetilde{\mathbb{Q}}_{\check{\alpha},+1}^{\enskip\check{a}}\otimes1+1\otimes\widetilde{\mathbb{Q}}_{\check{\alpha},+1}^{\enskip\check{a}}-\frac{1}{2}\overline{\mathbb{Q}}_{\check{\alpha}}^{\enskip\check{c}}\otimes\mathbb{R}_{\check{c}}^{\enskip\check{b}}+\frac{1}{2}\overline{\mathbb{R}}_{\check{c}}^{\enskip\check{a}}\otimes\mathbb{Q}_{\check{\alpha}}^{\enskip\check{c}}\nonumber \\
 & \qquad-\frac{1}{2}\overline{\mathbb{Q}}_{\check{\gamma}}^{\enskip\check{a}}\otimes\mathbb{L}_{\check{\alpha}}^{\enskip\check{\gamma}}+\frac{1}{2}\overline{\mathbb{L}}_{\check{\alpha}}^{\enskip\check{\gamma}}\otimes\mathbb{Q}_{\check{\gamma}}^{\enskip\check{a}}+\frac{1}{4}\overline{\mathbb{H}}\otimes\mathbb{Q}_{\check{\alpha}}^{\enskip\check{a}}-\frac{1}{4}\overline{\mathbb{Q}}_{\check{\alpha}}^{\enskip\check{a}}\otimes\mathbb{H},\nonumber \\
\Delta\widetilde{\mathbb{Q}}_{\check{\alpha},-1}^{\enskip\check{a}} & =\widetilde{\mathbb{Q}}_{\check{\alpha},-1}^{\enskip\check{a}}\otimes1+1\otimes\widetilde{\mathbb{Q}}_{\check{\alpha},-1}^{\enskip\check{a}}-\frac{1}{2}\varepsilon_{\check{\alpha}\check{\gamma}}\,\varepsilon^{\check{a}\check{d}}\,\overline{\mathbb{C}}\otimes\mathbb{G}_{\check{d}}^{\enskip\check{\gamma}}+\frac{1}{2}\varepsilon_{\check{\alpha}\check{\gamma}}\,\varepsilon^{\check{a}\check{d}}\,\overline{\mathbb{G}}_{\check{d}}^{\enskip\check{\gamma}}\otimes\mathbb{C},\nonumber \\
\Delta\widetilde{\mathbb{G}}_{\check{a},+1}^{\enskip\check{\alpha}} & =\widetilde{\mathbb{G}}_{\check{a},+1}^{\enskip\check{\alpha}}\otimes1+1\otimes\widetilde{\mathbb{G}}_{\check{a},+1}^{\enskip\check{\alpha}}+\frac{1}{2}\overline{\mathbb{G}}_{\check{c}}^{\enskip\check{\alpha}}\otimes\mathbb{R}_{\check{a}}^{\enskip\check{c}}-\frac{1}{2}\overline{\mathbb{R}}_{\check{a}}^{\enskip\check{c}}\otimes\mathbb{G}_{\check{c}}^{\enskip\check{\alpha}}\nonumber \\
 & \qquad+\frac{1}{2}\overline{\mathbb{G}}_{\check{a}}^{\enskip\check{\gamma}}\otimes\mathbb{L}_{\check{\gamma}}^{\enskip\check{\alpha}}-\frac{1}{2}\overline{\mathbb{L}}_{\check{\gamma}}^{\enskip\check{\alpha}}\otimes\mathbb{G}_{\check{a}}^{\enskip\check{\gamma}}-\frac{1}{4}\overline{\mathbb{H}}\otimes\mathbb{G}_{\check{a}}^{\enskip\check{\alpha}}+\frac{1}{4}\overline{\mathbb{G}}_{\check{a}}^{\enskip\check{\alpha}}\otimes\mathbb{H},\nonumber \\
\Delta\widetilde{\mathbb{G}}_{\check{a},-1}^{\enskip\check{\alpha}} & =\widetilde{\mathbb{G}}_{\check{a},-1}^{\enskip\check{\alpha}}\otimes1+1\otimes\widetilde{\mathbb{G}}_{\check{a},-1}^{\enskip\check{\alpha}}+\frac{1}{2}\varepsilon_{\check{a}\check{c}}\,\varepsilon^{\check{\alpha}\check{\gamma}}\,\overline{\mathbb{C}}^{\dagger}\otimes\mathbb{Q}_{\check{\gamma}}^{\enskip\check{c}}-\frac{1}{2}\varepsilon_{\check{a}\check{c}}\,\varepsilon^{\check{\alpha}\check{\gamma}}\,\overline{\mathbb{Q}}_{\check{\gamma}}^{\enskip\check{c}}\otimes\mathbb{C}^{\dagger}.\label{QG_D5}
\end{align}
This is the general structure of the secret symmetries for the reflection from $D5$-brane. 
The definitions of $\overline{\mathbb{C}}$, $\overline{\mathbb{C}}^\dagger$ and 
$\overline{\mathbb{H}}$ need to be developed a little further (see \cite{MR3} for complete details).
Two inequivalent orientations of the $D5$-brane, horizontal and vertical, that look 
rather different in the scattering theory were considered in \cite{CY,CRY}. 
Thus we will consider the explicit realization of the secret symmetries 
(\ref{B_D5}) and (\ref{QG_D5}) for both orientations separately.\\

In the case of reflection from the {\em horizontal} $D5$-brane ($\kappa=-1$),
the boundary is a singlet; thus neglecting the irrelevant terms in (\ref{B_D5}) and (\ref{QG_D5}) 
and with the help of the Lie algebra the remaining parts may be simplified to
\begin{equation}
\Delta\widetilde{\mathbb{B}}=\Bigl(\widehat{\mathbb{B}}\circ1-1\circ\widehat{\mathbb{B}}-\frac{1}{2}\bigl(\mathbb{Q}_{\alpha}^{\enskip a}\circ\mathbb{G}_{a}^{\enskip\alpha}+\mathbb{G}_{a}^{\enskip\alpha}\circ\mathbb{Q}_{\alpha}^{\enskip a}\bigr)\Bigr)\otimes1,\label{B_D5h}
\end{equation}
and
\begin{align}
\Delta\widetilde{\mathbb{Q}}_{\alpha,+1}^{\enskip a} & =\Bigl(\mathbb{Q}_{\alpha,+1}^{\enskip a}\circ1-1\circ\mathbb{Q}_{a,+1}^{\enskip\alpha}-\frac{1}{2}\mathbb{L}_{\alpha}^{\enskip\gamma}\circ\mathbb{Q}_{\gamma}^{\enskip a}+\frac{1}{2}\mathbb{Q}_{\gamma}^{\enskip a}\circ\mathbb{L}_{\alpha}^{\enskip\gamma}\nonumber \\
 & \qquad-\frac{1}{2}\mathbb{R}_{c}^{\enskip a}\circ\mathbb{Q}_{\alpha}^{\enskip c}+\frac{1}{2}\mathbb{Q}_{\alpha}^{\enskip c}\circ\mathbb{R}_{c}^{\enskip a}-\frac{1}{4}\mathbb{H}\circ\mathbb{Q}_{\alpha}^{\enskip a}+\frac{1}{4}\mathbb{Q}_{\alpha}^{\enskip a}\circ\mathbb{H}\Bigr)\otimes1,\nonumber \\
\Delta\widetilde{\mathbb{Q}}_{\alpha,-1}^{\enskip a} & =\Bigl(\mathbb{Q}_{\alpha,-1}^{\enskip a}\circ1-1\circ\mathbb{Q}_{a,-1}^{\enskip\alpha}-\frac{1}{2}\varepsilon_{\alpha\gamma}\,\varepsilon^{ac}\,\mathbb{G}_{c}^{\enskip\gamma}\circ\mathbb{C}+\frac{1}{2}\varepsilon_{\alpha\gamma}\,\varepsilon^{ac}\,\mathbb{C}\circ\mathbb{G}_{c}^{\enskip\gamma}\Bigr)\otimes1,\nonumber \\
\Delta\widetilde{\mathbb{G}}_{a,+1}^{\enskip\alpha} & =\Bigl(\mathbb{G}_{a,+1}^{\enskip\alpha}\circ1-1\circ\mathbb{G}_{a,+1}^{\enskip\alpha}+\frac{1}{2}\mathbb{L}_{\gamma}^{\enskip\alpha}\circ\mathbb{G}_{a}^{\enskip\gamma}-\frac{1}{2}\mathbb{G}_{a}^{\enskip\gamma}\circ\mathbb{L}_{\gamma}^{\enskip\alpha}\nonumber \\
 & \qquad+\frac{1}{2}\mathbb{R}_{a}^{\enskip c}\circ\mathbb{G}_{c}^{\enskip\alpha}-\frac{1}{2}\mathbb{G}_{c}^{\enskip\alpha}\circ\mathbb{R}_{a}^{\enskip c}+\frac{1}{4}\mathbb{H}\circ\mathbb{G}_{a}^{\enskip\alpha}-\frac{1}{4}\mathbb{G}_{a}^{\enskip\alpha}\circ\mathbb{H}\Bigr)\otimes1,\nonumber\\
\Delta\widetilde{\mathbb{G}}_{a,-1}^{\enskip\alpha} & =\Bigl(\mathbb{G}_{a,-1}^{\enskip\alpha}\circ1-1\circ\mathbb{G}_{a,-1}^{\enskip\alpha}+\frac{1}{2}\varepsilon_{ac}\,\varepsilon^{\alpha\gamma}\,\mathbb{Q}_{\gamma}^{\enskip c}\circ\mathbb{C}^{\dagger}-\frac{1}{2}\varepsilon_{ac}\,\varepsilon^{\alpha\gamma}\,\mathbb{C}^{\dagger}\circ\mathbb{Q}_{\gamma}^{\enskip c}\Bigr)\otimes1,\label{QG_D5h}
\end{align}
here `$\circ$' describes the tensor product of `left' and `right'
representations (hereafter `reps') of the bulk magnon and the usual tensor 
product `$\otimes$' separates the bulk and boundary reps. 
The central charges in this picture act on the bulk
states as $\mathbb{C}:=\mathbb{C}\circ1+1\circ\mathbb{C}$, 
$\,\overline{\mathbb{C}}:=\mathbb{C}\circ1-1\circ\mathbb{C}$
and analogously for $\mathbb{C}^{\dagger}$, $\mathbb{H}$. Note that
the secret charges (\ref{B_D5h}) and (\ref{QG_D5h}) effectively
differ from (\ref{secretB}) and (\ref{secretQG}) by a minus sign
only (see \cite{MR3} for the details on this similarity). 
We have checked that these charges commute with the reflection 
matrix $\mathcal{K}^{h}$ found in \cite{CRY}, and thus are secret 
symmetries of the horizontal $D5$-brane.\\

In the case of reflection from the {\em vertical} $D5$-brane ($\kappa=-i$),
the boundary carries a field multiplet transforming in the vector representation 
of the boundary algebra thus the non-local terms
in (\ref{B_D5}) and (\ref{QG_D5}) may no longer be neglected. Nevertheless
the general expressions may be casted in a quite transparent form, as
\begin{align}
\Delta\widetilde{\mathbb{B}} & =\Bigl(\widehat{\mathbb{B}}\circ1-1\circ\widehat{\mathbb{B}}-\frac{1}{2}\bigl(\mathbb{Q}_{\alpha}^{\enskip a}\circ\mathbb{G}_{a}^{\enskip\alpha}+\mathbb{G}_{a}^{\enskip\alpha}\circ\mathbb{Q}_{\alpha}^{\enskip a}\bigr)\Bigr)\otimes1\nonumber \\
 & \qquad-\frac{1}{2}(\mathbb{Q}_{\alpha}^{\enskip a}\circ1-1\circ\mathbb{Q}_{\alpha}^{\enskip a})\otimes\mathbb{G}_{a}^{\enskip\alpha}-\frac{1}{2}(\mathbb{G}_{a}^{\enskip\alpha}\circ1-1\circ\mathbb{G}_{a}^{\enskip\alpha})\otimes\mathbb{Q}_{\alpha}^{\enskip a},
\end{align}
and
\begin{align}
\Delta\widetilde{\mathbb{Q}}_{\alpha,+1}^{\enskip a} & =\Bigl(\mathbb{Q}_{\alpha,+1}^{\enskip a}\circ1-1\circ\mathbb{Q}_{a,+1}^{\enskip\alpha}-\frac{1}{2}\mathbb{L}_{\alpha}^{\enskip\gamma}\circ\mathbb{Q}_{\gamma}^{\enskip a}+\frac{1}{2}\mathbb{Q}_{\gamma}^{\enskip a}\circ\mathbb{L}_{\alpha}^{\enskip\gamma}\nonumber \\
 & \qquad-\frac{1}{2}\mathbb{R}_{c}^{\enskip a}\circ\mathbb{Q}_{\alpha}^{\enskip c}+\frac{1}{2}\mathbb{Q}_{\alpha}^{\enskip c}\circ\mathbb{R}_{c}^{\enskip a}-\frac{1}{4}\mathbb{H}\circ\mathbb{Q}_{\alpha}^{\enskip a}+\frac{1}{4}\mathbb{Q}_{\alpha}^{\enskip a}\circ\mathbb{H}\Bigr)\otimes1,\nonumber \\
 & \qquad-\frac{1}{2}(\mathbb{L}_{\alpha}^{\enskip\gamma}\circ1-1\circ\mathbb{L}_{\alpha}^{\enskip\gamma})\otimes\mathbb{Q}_{\gamma}^{\enskip a}+\frac{1}{2}(\mathbb{Q}_{\gamma}^{\enskip a}\circ1-1\circ\mathbb{Q}_{\gamma}^{\enskip a})\otimes\mathbb{L}_{\alpha}^{\enskip\gamma}\nonumber \\
 & \qquad-\frac{1}{2}(\mathbb{R}_{c}^{\enskip a}\circ1-1\circ\mathbb{R}_{c}^{\enskip a})\otimes\mathbb{Q}_{\alpha}^{\enskip c}+\frac{1}{2}(\mathbb{Q}_{\alpha}^{\enskip c}\circ1-1\circ\mathbb{Q}_{\alpha}^{\enskip c})\otimes\mathbb{R}_{c}^{\enskip a}\nonumber \\
 & \qquad-\frac{1}{4}(\mathbb{H}\circ1-1\circ\mathbb{H})\otimes\mathbb{Q}_{\alpha}^{\enskip a}+\frac{1}{4}(\mathbb{Q}_{\alpha}^{\enskip a}\circ1-1\circ\mathbb{Q}_{\alpha}^{\enskip a})\otimes\mathbb{H},\nonumber 
\end{align}
\begin{align}
\Delta\widetilde{\mathbb{Q}}_{\alpha,-1}^{\enskip a} & =\Bigl(\mathbb{Q}_{\alpha,-1}^{\enskip a}\circ1-1\circ\mathbb{Q}_{a,-1}^{\enskip\alpha}-\frac{1}{2}\varepsilon_{\alpha\gamma}\,\varepsilon^{ac}\,\mathbb{G}_{c}^{\enskip\gamma}\circ\mathbb{C}+\frac{1}{2}\varepsilon_{\alpha\gamma}\,\varepsilon^{ac}\,\mathbb{C}\circ\mathbb{G}_{c}^{\enskip\gamma}\Bigr)\otimes1\nonumber \\
 & \qquad-\frac{1}{2}\varepsilon_{\alpha\gamma}\,\varepsilon^{ac}(\mathbb{G}_{c}^{\enskip\gamma}\circ1-1\circ\mathbb{G}_{c}^{\enskip\gamma})\otimes\mathbb{C}+\frac{1}{2}\varepsilon_{\alpha\gamma}\,\varepsilon^{ac}(\mathbb{C}\circ1-1\circ\mathbb{C})\otimes\mathbb{G}_{c}^{\enskip\gamma},\nonumber \\
\Delta\widetilde{\mathbb{G}}_{a,+1}^{\enskip\alpha} & =\Bigl(\mathbb{G}_{a,+1}^{\enskip\alpha}\circ1-1\circ\mathbb{G}_{a,+1}^{\enskip\alpha}+\frac{1}{2}\mathbb{L}_{\gamma}^{\enskip\alpha}\circ\mathbb{G}_{a}^{\enskip\gamma}-\frac{1}{2}\mathbb{G}_{a}^{\enskip\gamma}\circ\mathbb{L}_{\gamma}^{\enskip\alpha}\nonumber \\
 & \qquad+\frac{1}{2}\mathbb{R}_{a}^{\enskip c}\circ\mathbb{G}_{c}^{\enskip\alpha}-\frac{1}{2}\mathbb{G}_{c}^{\enskip\alpha}\circ\mathbb{R}_{a}^{\enskip c}+\frac{1}{4}\mathbb{H}\circ\mathbb{G}_{a}^{\enskip\alpha}-\frac{1}{4}\mathbb{G}_{a}^{\enskip\alpha}\circ\mathbb{H}\Bigr)\otimes1\nonumber \\
 & \qquad+\frac{1}{2}(\mathbb{L}_{\gamma}^{\enskip\alpha}\circ1-1\circ\mathbb{L}_{\gamma}^{\enskip\alpha})\otimes\mathbb{G}_{a}^{\enskip\gamma}-\frac{1}{2}(\mathbb{G}_{a}^{\enskip\gamma}\circ1-1\circ\mathbb{G}_{a}^{\enskip\gamma})\otimes\mathbb{L}_{\gamma}^{\enskip\alpha}\nonumber \\
 & \qquad+\frac{1}{2}(\mathbb{R}_{a}^{\enskip c}\circ1-1\circ\mathbb{R}_{a}^{\enskip c})\otimes\mathbb{G}_{c}^{\enskip\alpha}-\frac{1}{2}(\mathbb{G}_{c}^{\enskip\alpha}\circ1-1\circ\mathbb{G}_{c}^{\enskip\alpha})\otimes\mathbb{R}_{a}^{\enskip c}\nonumber \\
 & \qquad+\frac{1}{4}(\mathbb{H}\circ1-1\circ\mathbb{H})\otimes\mathbb{G}_{a}^{\enskip\alpha}-\frac{1}{4}(\mathbb{G}_{a}^{\enskip\alpha}\circ1-1\circ\mathbb{G}_{a}^{\enskip\alpha})\otimes\mathbb{H},\nonumber \\
\Delta\widetilde{\mathbb{G}}_{a,-1}^{\enskip\alpha} & =\Bigl(\mathbb{G}_{a,-1}^{\enskip\alpha}\circ1-1\circ\mathbb{G}_{a,-1}^{\enskip\alpha}+\frac{1}{2}\varepsilon_{ac}\,\varepsilon^{\alpha\gamma}\,\mathbb{Q}_{\gamma}^{\enskip c}\circ\mathbb{C}^{\dagger}-\frac{1}{2}\varepsilon_{ac}\,\varepsilon^{\alpha\gamma}\,\mathbb{C}^{\dagger}\circ\mathbb{Q}_{\gamma}^{\enskip c}\Bigr)\otimes1\nonumber \\
 & \qquad+\frac{1}{2}\varepsilon_{ac}\,\varepsilon^{\alpha\gamma}(\mathbb{Q}_{\gamma}^{\enskip c}\circ1-1\circ\mathbb{Q}_{\gamma}^{\enskip c})\otimes\mathbb{C}^{\dagger}-\frac{1}{2}\varepsilon_{ac}\,\varepsilon^{\alpha\gamma}(\mathbb{C}^{\dagger}\circ1-1\circ\mathbb{C}^{\dagger})\otimes\mathbb{Q}_{\gamma}^{\enskip c}.
\end{align}
Once again we have checked that these new secret charges commute with the
reflection matrix $\mathcal{K}^{v}$ found in \cite{CRY} and 
the achiral reflection matrix found in \cite{MR2}, 
thus are the secret symmetries of the reflection from
the vertical $D5$-brane.


\section{Conclusions}

In this letter we have presented the secret symmetries appearing in the 
twisted boundary Yangians of the AdS/CFT reflection matrices. They originate 
from the known secret symmetries of the
AdS/CFT $S$-matrix and are generated by the secret charge $\Delta\widehat{\mathbb{B}}$.
These symmetries are called `secret' as they have no Lie algebra analogs, except for 
the $\Delta\widehat{\mathbb{B}}$ charge itself. However, even in this case, the corresponding 
Lie algebra charge $\mathbb{B}$, which acts as a helicity operator, 
is not a symmetry of the $S$-matrix.

We found that the twisted boundary Yangian of the $Y=0$ maximal giant
graviton enjoys the twisted secret symmetries $\Delta\widetilde{\mathbb{Q}}_{\alpha,\pm1}^{\enskip2}$
and $\Delta\widetilde{\mathbb{G}}_{2,\pm1}^{\enskip\alpha}$, but there
is no twisted partner of $\Delta\widehat{\mathbb{B}}$. The twisted secret
symmetry $\Delta\widetilde{\mathbb{B}}$ is not necessary for the $\Delta\widetilde{\mathbb{Q}}_{\alpha,\pm1}^{\enskip2}$
and $\Delta\widetilde{\mathbb{G}}_{2,\pm1}^{\enskip\alpha}$ to be present,
because in contrast to the Yangian of the $S$-matrix,
these charges are acquired from $\Delta\mathbb{Q}_{\alpha,\pm1}^{\enskip2}$
and $\Delta\mathbb{G}_{2,\pm1}^{\enskip\alpha}$, but not from $\Delta\widetilde{\mathbb{B}}$.
A plausible explanation why $\Delta\widetilde{\mathbb{B}}$ is not
a symmetry is because the fundamental reflection matrix is diagonal
and respects the helicity operator $\mathbb{B}$; hence the twisted Yangian partner of
$\widehat{\mathbb{B}}$ should not be a symmetry. The higher order (bound-state)
reflection matrices are of non-diagonal form and helicity symmetry
is broken, nevertheless it is easy to see that $\Delta\widetilde{\mathbb{B}}$
would not be a symmetry at any bound-state number.

Similarly, the mirror model of the $Y=0$ maximal giant graviton enjoys
secret symmetries $\Delta\widetilde{\mathbb{Q}}_{4,\pm1}^{\enskip a}$
and $\Delta\widetilde{\mathbb{G}}_{a,\pm1}^{\enskip4}$ and has no $\Delta\widetilde{\mathbb{B}}$.
However the reflection matrices are diagonal at all orders; thus helicity
charge $\mathbb{B}$ is a proper symmetry. Nevertheless this is not an
interesting situation as the reflection is trivial in the sense that at 
all orders it can be constrained by the Lie algebra alone and Yangian
symmetry is redundant.

The twisted Yangian of the $D5$-brane is the most rich in secrets. It enjoys
a complete set of twisted secret symmetries, 
$\Delta\widetilde{\mathbb{Q}}_{\check{\alpha},\pm1}^{\enskip\check{a}}$,
$\Delta\widetilde{\mathbb{G}}_{\check{a},\pm1}^{\enskip\check{\alpha}}$
and $\Delta\widetilde{\mathbb{B}}$ that may further be simplified to 
the specific expressions corresponding to the horizontal and vertical orientations 
of the $D5$-brane. 

Our results show that the twisted Yangians inherit most of the initial
Yangian properties, thus indicating that the secret symmetries are 
indeed a fundamental property of the Yangian of the centrally extended 
$\mathfrak{su}(2|2)$, as they appear in various physical models as 
AdS/CFT worldsheet scattering and reflection matrices, planar $\mathcal{N}=4$ 
super Yang-Mills MHV amplitudes, and perhaps shall be rediscovered 
in some other unexpected context.\\

{\bf Acknowledgements.} The author would like to thank Niall MacKay 
for many valuable discussions and a careful reading of the manuscript, 
Alessandro Torrielli for useful discussions and comments, and Burkhard Schwab 
for explanation of $\widehat{\mathfrak{B}}$ role in MHV amplitudes.
This work was supported by the UK EPSRC grant EP/H000054/1.


\newcommand{\nlin}[2]{\href{http://xxx.lanl.gov/abs/nlin/#2}{\tt nlin.#1/#2}}
\newcommand{\hepth}[1]{\href{http://xxx.lanl.gov/abs/hep-th/#1}{\tt hep-th/#1}}
\newcommand{\arXivid}[1]{\href{http://arxiv.org/abs/#1}{\tt arXiv:#1}}
\newcommand{\Math}[2]{\href{http://xxx.lanl.gov/abs/math.#1/#2}{\tt math.#1/#2}}

\end{document}